\documentclass[runningheads]{llncs}
\usepackage[square,comma,numbers,sort&compress,sectionbib]{natbib}
\usepackage{balance}
\usepackage[rightcaption]{sidecap}
\usepackage{xcolor}
\usepackage{latexsym}
\usepackage{pgfplots}
\pgfplotsset{compat=1.15}
\usepgfplotslibrary{
  statistics,
  colorbrewer,
  groupplots,
}
\usetikzlibrary{
  patterns,
  shapes.geometric,
  decorations.text,
  matrix,
  fit,
  backgrounds,
  positioning,
}
\tikzset{
  fignode/.style={
    outer sep=0.25em,
  }
}
\tikzset{
  framedfignode/.style={
    outer sep=0.25em,
    inner sep=0.5em,
    rounded corners,
    draw,
  }
}
\colorlet{plotColorNeutral}{gray}
\definecolor{plotColor1}{HTML}{f61a1c}
\definecolor{plotColor2}{HTML}{377eb8}
\definecolor{plotColor3}{HTML}{4daf4a}
\definecolor{plotColor4}{HTML}{984ea3}
\colorlet{plotColorNeutral*}{plotColorNeutral!40}
\colorlet{plotColor1*}{plotColor1!60}
\colorlet{plotColor2*}{plotColor2!60}
\colorlet{plotColor3*}{plotColor3!60}
\colorlet{plotColor4*}{plotColor4!60}
\pgfplotsset{
    colormap={greenred}{HTML=(4daf4a) HTML=(e41a1c)},
    colormap={redgreen}{HTML=(e41a1c) HTML=(4daf4a)}
}
\usepackage{colortbl}

\definecolor{blue}{RGB}{17,220,247}
\definecolor{purple}{RGB}{163,115,250}
\definecolor{caribbeangreen}{rgb}{0.0, 0.8, 0.6}

\definecolor{GREEN}{RGB}{84,130,53}

\newcommand{\colorg}{\cellcolor{gray!15}}
\usepackage{tikz}
\usetikzlibrary{positioning,shapes,decorations.pathmorphing}
\usepackage{tcolorbox}

\usepackage[utf8]{inputenc} 
\usepackage{microtype}      
\usepackage{xcolor}        
\usepackage{tabularx}
\usepackage{notes-alt} 
\usepackage{multirow}
\usepackage{longtable}
\usepackage{supertabular}
\usepackage{makecell}
\usepackage{graphicx}
\usepackage{hyperref}       
\usepackage{url}
\usepackage{multirow}
\usepackage{booktabs}       
\usepackage{amsfonts}       
\usepackage{nicefrac}       
\usepackage{amssymb}

\usepackage{svg}
\usepackage{enumitem}
\usepackage{caption}
\usepackage{subcaption}
\usepackage{ amsmath}
\usepackage[linesnumbered,ruled]{algorithm2e}
\usepackage{xcolor}

\definecolor{blue}{RGB}{17,220,247}
\definecolor{purple}{RGB}{163,115,250}

\newcommand{\speedup}[1]{\small ($\textcolor{green}{\blacktriangle}\textbf{#1})$}
\newcommand{\down}[1]{\small ($\textcolor{red}{\blacktriangledown}#1\%)$}

\usepackage{wrapfig}
\newcommand{\mpara}[1]{\medskip\noindent{\bf #1}}
\begin{document}
\mainmatter

% \title{Tackling Ambiguous Queries through Generative Query Rewriting}

\title{FlashCheck: Exploration of Efficient Evidence Retrieval for Fast Fact-Checking}

\titlerunning{FastCheck: Exploration of Efficient Evidence Retrieval for fast factChecking}
\titlerunning{FastCheck: Fast fact-checking through efficient evidence retreival}% abbreviated title (for running head)
% \author{Abhijit Anand\inst{1} \and Venktesh V\inst{2} \and Vinay Setty\inst{3}\orcidID{0000-0002-9777-6758} \and Avishek Anand\inst{2}}
% %
% \authorrunning{Anand et al.} % abbreviated author list (for running head)
% %
% %%%% list of authors for the TOC (use if author list has to be modified)
% \tocauthor{Abhijit Anand, Venktesh V, Vinay Setty, Avishek Anand}
% %
% \institute{L3S Research Institute, Germany \and Delft University of Technology, The Netherlands,  \and University of Stavanger, Norway \\
% \email{aanand@l3s.de, v.Viswanathan-1@tudelft.nl, vsetty@acm.org, Avishek.Anand@tudelft.nl}}
\author{%
Kevin Nanekhan\inst{1} \and
Venktesh V\inst{1} \and%\orcidID{0000-0001-9133-4978}
Erik Martin\inst{2,3}	 \and
Henrik Vatndal \inst{2,3} \and
Vinay Setty \inst{2,3} \and
Avishek Anand \inst{1}
}
\institute{%
Delft University of Technology \and
University of Stavanger \and Factiverse AI
}
\maketitle

\begin{abstract}

The advances in digital tools have led to the rampant spread of misinformation. While fact-checking aims to combat this, manual fact-checking is cumbersome and not scalable. It is essential for automated fact-checking to be efficient for aiding in combating misinformation in real-time and at the source. Fact-checking pipelines primarily comprise a knowledge retrieval component which extracts relevant knowledge to fact-check a claim from large knowledge sources like Wikipedia and a verification component. The existing works primarily focus on the fact-verification part rather than evidence retrieval from large data collections, which often face scalability issues for practical applications such as live fact-checking. In this study, we address this gap by exploring various methods for indexing a succinct set of factual statements from large collections like Wikipedia to enhance the retrieval phase of the fact-checking pipeline. We also explore the impact of vector quantization to further improve the efficiency of pipelines that employ dense retrieval approaches for first-stage retrieval. 
We study the efficiency and effectiveness of the approaches on fact-checking datasets such as HoVer and WiCE, leveraging Wikipedia as the knowledge source. We also evaluate the real-world utility of the efficient retrieval approaches by fact-checking 2024 presidential debate and also open source the collection of claims with corresponding labels identified in the debate. Through a combination of indexed facts together with Dense retrieval and Index compression, we achieve up to a \textbf{10.0x} speedup on CPUs and more than a \textbf{20.0x} speedup on GPUs compared to the classical fact-checking pipelines over large collections.

\end{abstract}

\vspace{-0.7cm}

\section{Introduction}
\label{sec:intro}

The rapid dissemination of disinformation and misinformation in the digital era has catastrophic consequences, impacting public opinion. Since manual fact-checking is not scalable and time-consuming, automated fact-checking approaches have been proposed~\cite{guo2022survey,yin-roth-2018-twowingos,read_twice,schuster2021vitamincrobustfact} to  combat misinformation. Automated fact-checking pipelines mimic the human fact-checking process by collecting multiple pieces of evidence for different aspects of the claim, followed by reasoning over evidence for claim verification~\cite{aly-vlachos-2022-natural,nakov2021automatedfactcheckingassistinghuman}.  Hence, these pipelines are usually comprised of a claim detection component, an evidence/knowledge retrieval component and a fact verification component \cite{thorne2018fever,programfc,jiang2020hover,kamoi2023wice}. Mitigating misinformation spread through platforms such as political debates and election campaigns, is time-critical due to its potential to proliferate faster, sway public opinion and its perceived reliability \cite{multimodal_spreads_faster,political_debates,multimodal_credibility}. Hence, it is essential that the verification pipeline is efficient to verify information at the source in real-time.

\noindent \textbf{The retrieval bottleneck}: One of the primary bottlenecks in existing fact-checking pipelines is the retrieval component employed for extracting information from large knowledge sources like Wikipedia. Existing approaches primarily employ sparse retrieval approaches followed by re-ranking or dense retrieval approaches \cite{samarinas-etal-2021-improving, jiang2020hover,schlichtkrull-etal-2021-joint} which perform search over a large index incurring huge computational, memory costs and high latency during inference. For instance, indexing and performing dense retrieval over the entire Wikipedia is prohibitively expensive with embedding index size being (\textbf{9.70 GiB}) and average retrieval latency per query (from WiCE) being \textbf{610 ms}. While several efficient sparse and dense Information Retrieval (IR) approaches exist \cite{leonhardt2023efficientneuralrankingusing,efficient_inverted_index,jegou2011pq,ge2014opq}, prior works in fact-checking primarily focus on fact verification components with very few works focusing on improving retrieval effectiveness \cite{zheng-etal-2024-evidence}. However, to the best of our knowledge, our work is the first to explore principled efficient IR approaches applied for fact-checking along with a case-study of its effectiveness for real-time fact-checking.

\noindent \textbf{Objectives}: We envision a fact-checking pipeline with an efficient retrieval component that has \textit{low memory footprint, computation requirements} and \textit{low latency} compared to existing systems. This would enable to run such systems in \textbf{low-resource settings}, making the technology more \textbf{accessible} to effectively combat misinformation at scale and possibly in real-time. 

Hence, in this work, we explore several approaches to extract and index succinct facts from large knowledge sources to aid in fact-checking claims. We observe that the corpus of knowledge sources such as Wikipedia could be compressed to \textbf{59\%} of the original information, optimizing sparse retrieval. Further, we also optimize dense retrieval by exploring vector quantization approaches for compressing the vector index of extracted factual statements which brings down the embeddings index size to \textbf{672.95 MiB} from 9.70 GiB. We observe that this approach leads to speedups of up to \textbf{10x} on the CPU and \textbf{20x} over the classic retrieve-rerank pipelines with the original corpus. The approaches we study also obviate the need for the snippet selection stage in fact-checking pipelines as it already indexes the succinct facts resulting in efficient inference.
 Through extensive experiments, we answer the following research questions:

\noindent \textbf{RQ1}: How does indexing factual statements for large collections affect information retrieval efficiency?

\noindent \textbf{RQ2}: How does indexing factual statements affect overall pipeline efficiency and downstream fact-checking performance?

\noindent \textbf{RQ3}: How does does index compression affect the efficiency of dense retrieval and overall fact-checking pipeline?

\noindent Following are our core contributions: 
\begin{itemize}
    \item We explore approaches aimed at efficiently identifying relevant text spans from large corpora pertaining to the claims being fact-checked, enabling efficient sparse retrieval and verification processes.
    
    \item We show the methods for indexing factual knowledge
    coupled with index compression further optimizes dense retrieval efficiency by reducing computational overhead compared to classical retrieve-rank-select pipelines without significant loss in performance.
    
    \item We deploy the efficient retrieval system for fact-checking the 2024 presidential debate in real-time and also open source the dataset of facts identified with corresponding human annotated labels.

\end{itemize}
\textbf{Reproducibility}: We open-source all our data and code under:
\texttt{\url{https://github.com/kevin-rn/Efficient-Fact-checking}}.

\vspace{-1em}
\section{Related Work}
\vspace{-1em}
The rapid proliferation of misinformation and disinformation has given rise to the development of automated fact-checking systems to combat them \cite{thorne2018fever,programfc,guo2022survey,v2024quantemprealworldopendomainbenchmark,questgen}. The automated fact-checking process involves three stages comprising \textit{claim detection}, which identifies salient spans to be fact-checked, 
followed by \textit{evidence retrieval} that focuses on identifying sources that support or refute claims and finally the fact-verification stage that uses the evidence collected to categorize the claims.

A significant challenge in verification includes source reliability \cite{guo2022survey}, hence fact-checking primarily relies on verified knowledge sources. Sources such as encyclopedias, policy documents, and scientific journals are common knowledge sources employed for retrieving information \cite{Lazarski2021nlpfact, thorne2018automated}. Recent advancements advocate a simplified two-step evidence retrieval approach: a context retriever selects a subset of passages that might contain the answer followed by a machine reader analyzing these passages to identify the correct answer. Initially, evidence retrieval relied on inverted indexes (e.g., TF-IDF, BM25) for keyword-based searches but these lack semantic understanding \cite{baranchuk2018revisiting, wei2022}. The advances in representation learning have led to the rise of dense retrieval where queries and documents are projected to a continuous vector space  \cite{karpukhin2020dense, zhao2022,Guo_2022}. While existing approaches employ brute force search in the vector space to retrieve relevant documents for the queries, this is not scalable for web-scale search \cite{bondarenko2021understanding, zhu2023survey, han2023comprehensive, wei2022,tas_b}. Compact vector representations are crucial for efficiency, despite potential noise-induced performance drops \cite{zhan2021jointly}. Efficient approaches like Approximate Nearest Neighbors (ANN) search in vector databases have emerged to reduce complexity and enhance similarity search accuracy \cite{han2023comprehensive, wei2022, zhao2023ann}. Early on hash-based and tree-based methods were used but faced limitations in large-scale databases and semantic features. Recent advancements make use of quantization techniques, such as Product Quantization\cite{jegou2011pq} and Optimized PQ (OPQ) \cite{ge2014opq}, to improve efficiency by vector dimension reduction with minimal performance loss.

Our contributions aim to explore efficient retrieval approaches to improve the efficiency of the fact-checking process, enhancing the practical applicability of these systems in real-world scenarios such as live fact-checking over large knowledge bases.

\section{Methodology}
\begin{figure*}[hbt!]
    \centering
    \includegraphics [width=1.0\textwidth]{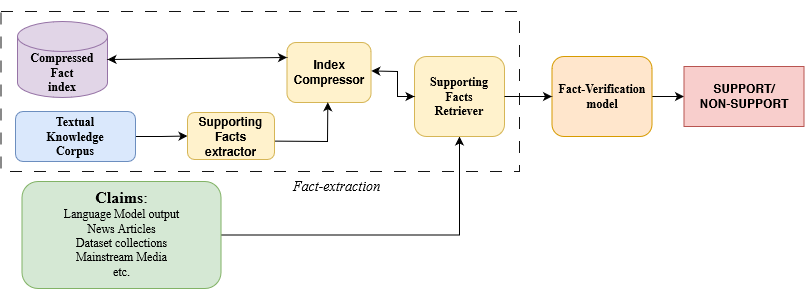}
    \caption{Comparison of Existing and Proposed Fact-Checking Pipelines}
    \label{fig:original_pipeline}
\end{figure*}

In this section, we describe our proposed improvements on how to enhance fact-checking efficiency by optimizing the retrieval of evidence from knowledge sources. This involves constructing a pruned index of succinct facts from large sources, like Wikipedia, coupled with vector quantization for index compression to improve retrieval latency and reduce resource consumption, as illustrated in \autoref{fig:original_pipeline}. 

\vspace{-2em}
\subsection{Corpus Compression through Extraction of Facts}\label{ssec:reranking}
 Large collections like Wikipedia are usually employed as knowledge sources for fact-checking \cite{jiang2020hover,kamoi2023wice}. While each Wikipedia article has detailed information on a topic, all the information is not factual, lacking citations and less informative for purposes of verifying claims. Hence, identifying useful factual information for fact-checking from such a large corpus and pruning other parts can significantly reduce pipeline runtime and save disk space for indexing. We employ different mechanisms to identify and store factual statements from Wikipedia corpus.
 
Since factual statements are precise statements that are verifiable, we posit that claim detection can help identify such succinct facts. Hence, we first employ a claim-detection model to identify significant (salient) sentences within the text, aiming to extract key information effectively. For experimental purposes, a model trained on a claim detection dataset such as Claimbuster is leveraged. We term this approach \textbf{Fact Extraction (FE)}

However, this approach could result in false positives leading to statements that do not contain factual information or aid in verifying claims. It would also result in false negatives as the statements the claim detection model fails to identify due to distribution mismatch would lead to incorrect verdicts. To circumvent this, we explore an alternative approach with a stricter criterion of retaining only verified factual statements on Wikipedia articles. We retain only those snippets that comprise citations following Wikipedia's \textit{verifiability} \footnote{https://en.wikipedia.org/wiki/Wikipedia:Verifiability} and \textit{citation} \footnote{https://en.wikipedia.org/wiki/Wikipedia:Citing\_sources} guidelines that the cited statements ensure verifiability of information conveyed. This method eliminates the need for training a specific model, providing a straightforward way to identify factual statements. We term this approach \textbf{Citation Extraction (CE)} in our experiments. However, it may miss contextually related sentences, such as those in Wikipedia, where consecutive sentences can be supported by a single citation, resulting in false negatives. 

To balance the tradeoffs in the above approaches, we focus on minimizing false negatives to capture all information that would aid in verifying claims. We propose to fuse the snippets identified by the claim detection-based approach and the citation extraction approach while pruning other information from the Wiki pages. While the problem of false positives from claim detection would persist, we posit that this can be minimized through retrieval and re-ranking approaches that optimize for recall.
The fusion of the former approaches aims to leverage the strengths of each approach: citations for reliable support and claim detection for comprehensive coverage. We term this approach \textbf{Fusion (Fu)} in our experiments.

While pruning the Wikipedia collection helps reduce memory requirements for storing and using the plaintext collection, it also helps optimize latency for sparse retrieval approaches like BM25. However, dense retrieval approaches provide better semantic relevance estimates than sparse retrieval and hence are employed in many fact-checking pipelines. However, dense retrieval is expensive due to index size comprising document vector representations and also has high latency due to brute force search approaches for retrieving relevant documents for a query. Hence, in this work, we explore an index compression mechanism that reduces index size, while ensuring optimal performance. We term this approach \textbf{Index Compression} in our experiments.

\subsection{Index Compression}
Existing fact-checking systems usually adopt sparse retrieval followed by re-ranking or dense retrieval approaches to capture better semantics for first-stage retrieval. However, dense retrieval using brute force search for web-scale data is not scalable. To enhance efficiency in Dense retrieval, we propose to compress the corpus embeddings index through the JPQ index compression algorithm \cite{zhan2021jointly}. Unlike existing approaches that treat training of encoders and learning compressed index separately, JPQ optimizes them jointly by employing a ranking-oriented loss produced by the encoders and the index. JPQ achieves an impressive compression ratio of \textit{4D/M} (where \textit{D} is vector dimensionality and \textit{M} is the number of codebooks (embedding sets)) \cite{zhan2021jointly}. This approach ensures efficient retrieval with a speedup ratio of $({D + \log n})/({M + \log n})$, where \textit{n} is the total number of documents, maintaining performance comparable to standard Dense Retrieval setups while maintaining a small memory footprint for the index. JPQ is based on Product Quantization (PQ) which quantizes the document embeddings $\vec{d}^\phi$ where $\phi$ denotes they are quantized. Quantization occurs by generating $M$ sets of embedding, where each set has $K$ embeddings of dimension $D/M$ called centroid embeddings $\vec{c}_{i,j}$. A quantized embedding for a document is constructed by selecting one embedding from each set and concatenating them. The matching between queries and quantized embeddings is performed using approximate score function $s^\phi(q, d) = \langle \vec{q},\vec{d}^\phi \rangle$. The search in PQ happens efficiently between query embeddings and quantized document embeddings by generating query sub-vectors and matching them to centroid embeddings.\\

The traditional dense retrieval models \cite{luan-etal-2021-sparse, zhan2020repbertcontextualizedtextembeddings} use the pairwise loss,
$l(s(q,d^+), \\ s(q,d^-))$ where $q$ is the query and $d^+ \in D_q^+$ indicates a positive document and $d^- \in D_q^+$ indicates a negative document with respect to the query. To adopt this loss for quantization setting the score function $s(q,d)$ has to be replaced by $s^\phi(q,d)$. Prior quantization approaches do not adopt a ranking-oriented loss and train the retrievers/dual-encoders independently of compression. To jointly train the quantization and retrieval approach end-end,
JPQ trains the query encoder and PQ index jointly for the complete optimization objective formulated as
\begin{equation}
\langle f^*, \{c_{i,j}\}^* \rangle = \arg\min_{f,\{c_{i,j}\}} \sum_{i}\sum_{d^+ \in D^{+}_i}\sum_{d^- \in D^{\phi-}_i} \ell(s^\phi(q, d^+), s^\phi(q, d^-))
\end{equation} 
, where $D^{\phi-}_i$ refers to the retrieved hard-negatives. Optimizing encoders to penalize hard-negatives is crucial to retrieval performance and JPQ obtains real-time hard-negatives by retrieving hard-negatives with respect to the current query from the quantized index being learned simultaneously thus optimizing for retrieval and quantization performance.

\section{Experimental Setup}
\label{experiments}

\subsection{Datasets}

\noindent \textbf{HoVer}\cite{jiang2020hover} consists of 26k claims requiring evidence from up to four English Wikipedia articles (2017 data dump) and was developed in multiple stages with the help of trained crowd-workers. The dataset is split into 18,171 labeled training examples (11,023 Supported and 7,148 Not-Supported), 4,000 labeled development examples (2,000 Supported and 2,000 Not-Supported), and 4,000 unlabeled test examples. 

\noindent \textbf{WiCE}\cite{kamoi2023wice} focuses on claims from Wikipedia articles. It uses the same base claims as SIDE \cite{petroni2022improving} and breaks down long claims into simpler sub-claims, enhancing annotation and entailment prediction. For our experiments, we used the original WiCE claims instead of sub-claims to retain context. We further adapted WiCE's three-way entailment to a binary scheme by relabelling the partially supported claims as unsupported for aligning with the HoVer dataset. The dataset is split into 1,260 labeled training examples (460 Supported and 800 Not-Supported), 349 labeled development examples (115 Supported and 234 Not-Supported, and 358 unlabeled test examples.

%It should be noted that the unsupported claims are likely to be true but lack specific evidence in the document corpus.

\noindent \textbf{Wikipedia corpus}: For HoVer, we used the processed 2017 English Wikipedia dump from the HotPotQA website, containing 5,486,211 articles. For WiCE, we used the latest available English Wikipedia dump from January 1, 2024, which includes 6,777,401 articles. Both dumps were processed using the HotPotQA fork\footnote{https://github.com/qipeng/wikiextractor} of the Wikiextractor tool\footnote{https://github.com/attardi/wikiextractor} to format the data into a structured folder system containing Bzip2 files.

\noindent \textbf{2024 Presidential debate:} We employ the fact-checking pipeline for the task of live fact-checking of the 2024 US presidential debate. We identified 281 claims in the debate with corresponding labels. We employ the 2024 Wikipedia dump as knowledge source for fact-checking the claims.

\vspace{-1.2em}
\subsection{Experiment Setup}
\noindent \textbf{Computational resources} 
The experiments were conducted on a dedicated server running Arch Linux, equipped with a 16-core 2nd Gen AMD EPYC™ 7302 processor, 256 GB RAM and two NVIDIA GeForce RTX 3090 GPUs.
%Our code relies mostly on Python 3.10.9 to utilise the latest models and tools, except for parts of the legacy HoVer pipeline code which relies on Python 3.7.16 to maintain stability. 

\noindent \textbf{Tools, Models and Hyperparameters} The Wikipedia dumps are processed using WikiExtractor, JobLib (parallelization) and SpaCy's `en\_core\_web\_lg' model\footnote{https://spacy.io/models/en} for sentence splitting instead of StanfordCoreNLP due to performance issues (processing large volumes of text in parallel). For \textbf{sparse retrieval}, we utilize BM25 from Elasticsearch\footnote{https://elasticsearch-py.readthedocs.io/en/v8.12.1/}. For \textbf{dense retrieval}, we employ \textit{`all-MiniLM-L6-v2'} model from Sentence Transformers\footnote{https://www.sbert.net/docs/pretrained\_models.html} to encode the query and all the documents. We employ ANN search using a flat index structure over the document embeddings leveraging FAISS library\footnote{https://github.com/facebookresearch/faiss}.
 For index compression JPQ, configured with \textit{K=256} codewords and \textit{M=96} codebooks, utilizes Roberta-based dual-encoders with OPQ for linear transformation and PQ for compression. Training involved AdamW optimizer with batch size 32, LambdaRank for pair-wise loss, and optimization settings specific to query and PQ learning rates. For claim detection, a pre-trained BERT model\footnote{https://huggingface.co/Nithiwat/bert-base\_claimbuster} fine-tuned on the ClaimBuster dataset is used. Other pre-trained BERT-base uncased models are employed in the pipeline, fine-tuned with a batch size of 16, a learning rate of 3e-5, and varying epochs (3 for Sentence Selection, 5 for other stages).
 %Model parameters like kr, kp, κp, and κs are adjusted based on memory constraints and performance on development claim datasets (e.g., kr=20, kp=5, κp=0.5, and κs=0.3).

\noindent \textbf{Performance and Efficiency Evaluation:} To evaluate end-end task performance, we employ weighted F1 score. For retrieval latency, we'll measure CPU and GPU implementations of FAISS and CPU for BM25. Metrics such as index size, creation time, retrieval time, total runtime, dataset size, and disk writes will be monitored using tools like psutil\footnote{https://psutil.readthedocs.io/en/latest/} and nvidia-ml-py3\footnote{https://github.com/nicolargo/nvidia-ml-py3}. Inference latency will be measured in milliseconds.

\section{Results}
\label{sec:results}
% In this section, we answer the research questions formulated in Section \ref{experiments}.

\subsection{Effectiveness of indexing succinct facts to improve information retrieval efficiency} To answer \textbf{RQ1}, we measure the memory and computational costs of fact-checking using full-Wikipedia compared to the pruned version proposed in this work. We first measure the index size on disk, measuring the raw JSON file size containing the article titles and texts, for each experiment setting. In \autoref{fig:disk_size}, we observe a significant reduction in disk space usage with HoVer having a reduction ranging from \textbf{44-55\%}, and WiCE from \textbf{44-57\%}. Additionally, the number of sentences stored in the index also decreases, with HoVer showing a reduction from 52-61\% and WiCE 52-59\%, indicating that at least half of the sentences are not helpful in claim verification.

\begin{table}[htb!]
\small
\centering
\footnotesize
\begin{tabular}{c c c c c}
\toprule
\small
\textbf{Method} & \textbf{Disk Size} & \textbf{Size reduction} & \textbf{\#Sentences} & \textbf{\% decrease}\\
\hline \hline
\multicolumn{1}{l}{\colorg\textbf{HoVer}} & \colorg& \colorg & \colorg & \colorg  \\
Full-Wiki  &  11.28 GiB & - & 94,914,378 & -   \\
Fact Extraction & 6.19 GiB & \down{45}& 45,894,704 & \down{52} \\
Citation Extraction & \textbf{5.07 GiB} & \down{\textbf{55} } & \textbf{36,886,889} & \down{\textbf{61}} \\
Fusion & 5.45 GiB & \down{52} & 39,842,574 & \down{{58}} \\

\hline
\multicolumn{1}{l}{\colorg \textbf{WiCE}} & \colorg &  \colorg & \colorg & \colorg \\
Full-Wiki & 15.28 GiB & - &  126,533,841 & -  \\
Fact Extraction & 8.56 GiB & \down{44} & 61,040,380 & \down{52}\\
Citation Extraction & \textbf{6.56 GiB} & \down{\textbf{57}} & \textbf{51,735,961} & \down{\textbf{59}}  \\
Fusion & 6.85 GiB & \down{55}& 54,070,295 & \down{57} \\

\hline
\end{tabular}
\caption{Comparison sizes for the corpora per experiment setting, consisting of English Wikipedia articles 2017 (HoVer) and 2024 (WiCE). Reduction is measured compared to the  Full-Wiki data setting. \down{} denotes a reduction in corpus size and number of sentence compared to Full-Wiki setting.}
\label{fig:disk_size}
\end{table}
\vspace{-2em}

 Following the reduction in disk size, a notable improvement in retrieval latency is evident, as demonstrated in \autoref{tab:bm25_latency}.
Regarding document retrieval latency (which encompasses both column values), there's an observed speedup ranging from approximately 1.5x (334 ms) to 1.6x (316 ms) compared to the original experimental setting for HoVer (495 ms). Similarly, in WiCE experiments, we witness a comparable speedup rate ranging from 1.4x (446 ms) to 1.6x (399 ms) compared to the original experimental setting (636 ms). This observation suggests that while the reduced text size contributes to efficient retrieval, it could further be improved.

\begin{table}[htb!]
\centering
\small
\footnotesize
% \vspace{-1cm}
\begin{tabular}{l c c c c}
\multirow{2}{*}{\makecell{\textbf{Methods}}} & \multirow{2}{*}{\textbf{Retrieval}} & \multirow{2}{*}{\makecell{\textbf{Total Latency}}} & \multirow{2}{*}{\makecell{\textbf{Speedup}}} \\
& \\
\hline
\multicolumn{1}{l}{\colorg\textit{HoVer}} & \colorg & \colorg & \colorg \\
 Full-Wiki & 426  ms & 659 ms & - \\
Fact Extraction & 257 ms  & 338  ms & 1.9x \\
Citation Extraction & 246 ms  &  327  ms & \speedup{2.0x}  \\
Fusion & 265 ms  & 345  ms & 1.9x  \\

\multicolumn{1}{l}{\colorg\textit{WiCE}} & \colorg & \colorg & \colorg \\
  Full-Wiki &  559 ms  &  831  ms & - \\
Fact Extraction & 372 ms  & 468   ms & 1.8x \\
Citation Extraction & 330 ms   & 419  ms & \speedup{2.0x} \\
Fusion & 347 ms & 436  ms & 1.9x \\
\hline
\end{tabular}
\caption{Retrieval and total latency for Sparse retrieval with Re-ranking. Speedup is compared with respect to the total latency of the Full-Wiki setup.}
\label{tab:bm25_latency}
\vspace{-2em}
\end{table}

% \input{tables/latency/faiss_latency}

% However, sparse retrieval does not capture semantic information and requires a costly re-ranking stage. While transitioning from Sparse to Dense retrieval may help improve the performance dense retrieval introduces additional computational costs as seen in Table \ref{tab:faiss_latency}. This is due to our use of FAISS utilising fixed-dimensionality vectors where despite varying article text lengths, the constant number of text embeddings minimizes impact on retrieval speed. However, dense retrieval libraries offer GPU support, which can yield substantial speedups compared to the CPU-based retrieval of BM25 and FAISS. GPU retrieval shows substantial speedups: 16.6-22.3x for HoVer and 17.9-20.2x for WiCE compared to CPU retrieval. Compared to BM25, GPU retrieval offers 16.0-21.5x speedup for HoVer and 18.7-20.5x speedup for WiCE. Thus, making Dense Retrieval a highly efficient and viable option over standard Sparse Retrieval with re-ranking.
%\vspace{-0.5em}

\noindent \textbf{Insight 1}: \textit{
Extraction of succinct facts reduces storage requirements and improves latency for Sparse retrieval while only leading to a minor loss in task performance.}
\vspace{-2em}
% %%%%%%%%%%%%%%%%%%%%%%%%%%%%%%%%%%%%%%%%%%%%%%%%%%%%%%%%%%%%%%%%%%%%%%%%%%%%%%%%%%%

\subsection{Effectiveness of pruned knowledge sources on overall pipeline efficiency and downstream fact-checking performance?}
To answer \textbf{RQ2}, we now shift focus to analyzing the inference time throughout the entire pipeline instead of solely the retrieval part. Extraction of just supporting facts not only has a improvement in the retrieval stage but also on the overall inference latency across the pipeline. For HoVer this being a 1.9-2.0x speedup and 1.8-2.0x for WiCE experiments. This improvement can be attributed to not only faster retrieval times but also the elimination of the Sentence Retrieval stage, which previously imposed significant latency overhead. 

%\input{tables/hover/hover_performance}
% \begin{figure}
%     \centering
%     \includegraphics[width=0.82\linewidth]{figs/graphs/hover_perf.png}
%     \caption{HoVer performance comparison}
%     \label{fig:enter-label}
% \end{figure}
% %\input{tables/wice/wice_performance}

% \begin{figure}[hbt!]
%     \centering
%     \includegraphics[width=0.85\linewidth]{figs/graphs/wice_perf.png}
%     \caption{WiCe performance comparison}
%     \label{fig:enter-label}
% \end{figure}

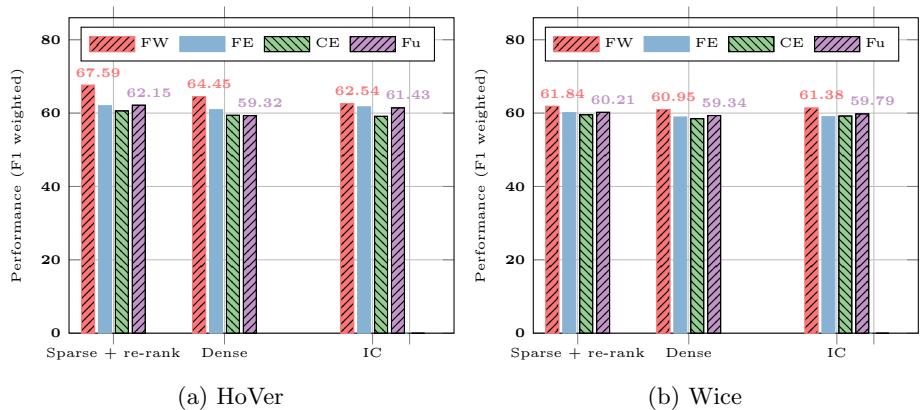
\begin{figure*}[hbt!]
    \begin{subfigure}{.5\textwidth}
        \begin{tikzpicture}
\edef\mylst{"67.59","64.45","62.54"}
\edef\explora{"62.15","59.32","61.43"}

    \begin{axis}[
            ybar=1.5pt,
            width=6.7cm,
            bar width=0.35,
            every axis plot/.append style={fill},
            grid=major,
            xtick={1, 4, 8,9,11},
            xticklabels={Sparse + re-rank, Dense, IC},
            ylabel style = {font=\tiny},
        yticklabel style = {font=\boldmath \tiny,xshift=0.05ex},
        xticklabel style ={font=\tiny,yshift=0.5ex},
            ylabel={Performance (F1 weighted)},
            enlarge x limits=0.15,
            ymin=0,
            ymax=86,
            legend style ={font=\tiny,yshift=0.05ex},
            area legend,
            nodes near coords style={font=\tiny,align=center,text width=1em},
            legend entries={FW, FE, CE, Fu},
            legend cell align={left},
            legend pos=north west,
            legend columns=-1,
            legend style={/tikz/every even column/.append style={column sep=0.06cm}},
        ]
        \addplot+[
            ybar,
            plotColor1*,
            postaction={
                    pattern=north east lines
                },
                    nodes near coords=\pgfmathsetmacro{\mystring}{{\mylst}[\coordindex]}\textbf{\mystring},
        ] plot coordinates {
                (1,67.59)
                (4,64.45)
                (8,62.54)
            };
        \addplot+[
            ybar,
            plotColor2*,
        ] plot coordinates {
                (1,62.02)
                (4,60.94)
                (8,61.71)
                (9,0)
            };

                    \addplot+[
            ybar,
            plotColor3*,
            draw=black,
    nodes near coords align={vertical},
            postaction={
                    pattern=north west lines
                },
        ] plot coordinates {
                (1,60.61)
                (4,59.40)
                (8,59.12)
                (9,0)
            };
             \addplot+[
            ybar,
            plotColor4*,
            draw=black,
            postaction={
                    pattern=north east lines
                },
            nodes near coords=\pgfmathsetmacro{\mystring}{{\explora}[\coordindex]}\textbf{\mystring},
        ] plot coordinates {
                (1,62.15)
                (4,59.32)
                (8,61.43)
            };
    \end{axis}
\end{tikzpicture}
\subcaption{HoVer}
    \end{subfigure}
        \begin{subfigure}{.5\textwidth}
    \begin{tikzpicture}
\edef\mylst{"61.84","60.95","61.38"}
\edef\explora{"60.21","59.34","59.79"}

    \begin{axis}[
            ybar=1.5pt,
            width=6.7cm,
            bar width=0.35,
            every axis plot/.append style={fill},
            grid=major,
            xtick={1, 4, 8,9,11},
            xticklabels={Sparse + re-rank, Dense, IC},
            ylabel style = {font=\tiny},
        yticklabel style = {font=\boldmath \tiny,xshift=0.05ex},
        xticklabel style ={font=\tiny,yshift=0.5ex},
            ylabel={Performance (F1 weighted)},
            enlarge x limits=0.15,
            ymin=0,
            ymax=86,
            legend style ={font=\tiny,yshift=0.05ex},
            area legend,
            nodes near coords style={font=\tiny,align=center,text width=1em},
            legend entries={FW, FE, CE, Fu},
            legend cell align={left},
            legend pos=north west,
            legend columns=-1,
            legend style={/tikz/every even column/.append style={column sep=0.06cm}},
        ]
        \addplot+[
            ybar,
            plotColor1*,
            postaction={
                    pattern=north east lines
                },
                    nodes near coords=\pgfmathsetmacro{\mystring}{{\mylst}[\coordindex]}\textbf{\mystring},
        ] plot coordinates {
                (1,61.84)
                (4,60.95)
                (8,61.38)
            };
        \addplot+[
            ybar,
            plotColor2*,
        ] plot coordinates {
                (1,60.12)
                (4,58.94)
                (8,59.02)
                (9,0)
            };

                    \addplot+[
            ybar,
            plotColor3*,
            draw=black,
    nodes near coords align={vertical},
            postaction={
                    pattern=north west lines
                },
        ] plot coordinates {
                (1,59.56)
                (4,58.48)
                (8,59.21)
                (9,0)
            };
             \addplot+[
            ybar,
            plotColor4*,
            draw=black,
            postaction={
                    pattern=north east lines
                },
            nodes near coords=\pgfmathsetmacro{\mystring}{{\explora}[\coordindex]}\textbf{\mystring},
        ] plot coordinates {
                (1,60.21)
                (4,59.34)
                (8,59.79)
            };
    \end{axis}
\end{tikzpicture}
    \subcaption{Wice}

    \end{subfigure}
    \caption{HoVer and WiCe task performance (FW- Full-Wiki, FE - Fact Extraction, IC- Index Compression, CE - Citation Extraction, Fu - Fusion)}
    \label{fig:performance_plot}
    \end{figure*}

\begin{figure*}[hbt!]
    \begin{subfigure}{.5\textwidth}
        \begin{tikzpicture}
\edef\mylst{"67.59","64.45","62.54"}
\edef\explora{"62.15","59.32","61.43"}

    \begin{axis}[
            ybar=1.5pt,
            width=6.2cm,
            bar width=0.35,
            every axis plot/.append style={fill},
            grid=major,
            xtick={1, 4, 8,9,11},
            xticklabels={Sparse + re-rank, Dense, IC},
            ylabel style = {font=\tiny},
        yticklabel style = {font=\boldmath \tiny,xshift=0.05ex},
        xticklabel style ={font=\tiny,yshift=0.5ex},
            ylabel={Recall@10},
            enlarge x limits=0.15,
            ymin=0,
            ymax=0.5,
            legend style ={font=\tiny,yshift=0.05ex},
            area legend,
            nodes near coords style={font=\tiny,align=center,text width=1em},
            legend entries={FW, FE, CE, Fu},
            legend cell align={left},
            legend pos=north west,
            legend columns=-1,
            legend style={/tikz/every even column/.append style={column sep=0.06cm}},
        ]
        \addplot+[
            ybar,
            plotColor1*,
            postaction={
                    pattern=north east lines
                },
        ] plot coordinates {
                (1,0.136)
                (4,0.123)
                (8,0.098)
            };
        \addplot+[
            ybar,
            plotColor2*,
        ] plot coordinates {
                (1,0.105)
                (4,0.094)
                (8,0.098)
                (9,0)
            };

                    \addplot+[
            ybar,
            plotColor3*,
            draw=black,
    nodes near coords align={vertical},
            postaction={
                    pattern=north west lines
                },
        ] plot coordinates {
                (1,0.126)
                (4,0.143)
                (8,0.096)
                (9,0)
            };
             \addplot+[
            ybar,
            plotColor4*,
            draw=black,
        ] plot coordinates {
                (1,0.124)
                (4,0.141)
                (8,0.097)
            };
    \end{axis}
\end{tikzpicture}
\subcaption{WiCE (nDCG@10)}
    \end{subfigure}
        \begin{subfigure}{.5\textwidth}
    \begin{tikzpicture}
\edef\mylst{"67.59","64.45","62.54"}
\edef\explora{"62.15","59.32","61.43"}

    \begin{axis}[
            ybar=1.5pt,
            width=6.4cm,
            bar width=0.35,
            every axis plot/.append style={fill},
            grid=major,
            xtick={1, 4, 8,9,11},
            xticklabels={Sparse + re-rank, Dense, IC},
            ylabel style = {font=\tiny},
        yticklabel style = {font=\boldmath \tiny,xshift=0.05ex},
        xticklabel style ={font=\tiny,yshift=0.5ex},
            ylabel={Recall@10},
            enlarge x limits=0.15,
            ymin=0,
            ymax=0.5,
            legend style ={font=\tiny,yshift=0.05ex},
            area legend,
            nodes near coords style={font=\tiny,align=center,text width=1em},
            legend entries={FW, FE, CE, Fu},
            legend cell align={left},
            legend pos=north west,
            legend columns=-1,
            legend style={/tikz/every even column/.append style={column sep=0.06cm}},
        ]
        \addplot+[
            ybar,
            plotColor1*,
        ] plot coordinates {
                (1,0.309)
                (4,0.195)
                (8,0.160)
            };
        \addplot+[
            ybar,
            plotColor2*,
        ] plot coordinates {
                (1,0.241)
                (4,0.166)
                (8,0.163)
                (9,0)
            };

                    \addplot+[
            ybar,
            plotColor3*,
            draw=black,
    nodes near coords align={vertical},
            postaction={
                    pattern=north west lines
                },
        ] plot coordinates {
                (1,0.295)
                (4,0.226)
                (8,0.174)
                (9,0)
            };
             \addplot+[
            ybar,
            plotColor4*,
            draw=black,
        ] plot coordinates {
                (1,0.286)
                (4,0.223)
                (8,0.160)
            };
    \end{axis}
\end{tikzpicture}
    \subcaption{WiCE (Recall@10)}

    \end{subfigure}
    \caption{Retrieval performance comparison}
    \label{fig:retrieval_perf}
    \end{figure*}
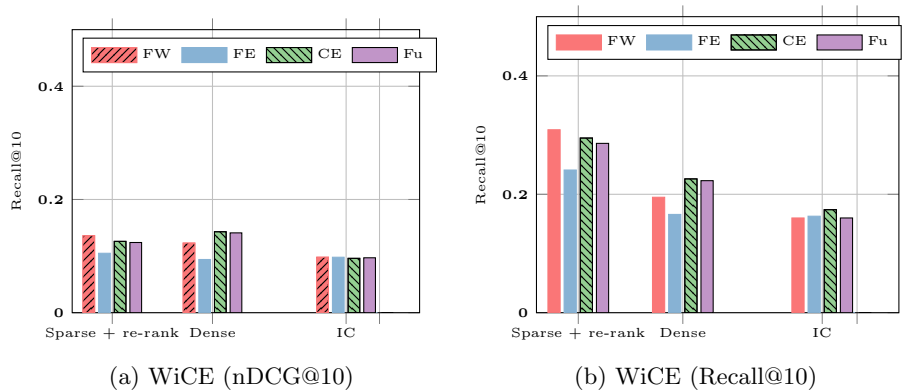
The evaluation of Sparse and Dense Retrieval setups in HoVer and WiCE experiments reveals that Sparse Retrieval, particularly fact extraction (FE) and Fusion approaches, maintains performance closest to the Full-Wiki setup as measured by weighted F1 in Figure \ref{fig:performance_plot}, while citation extraction has a larger drop in performance. Most notably, the Fusion method compared to the other methods has relatively high scores, underscoring the importance of combining supporting facts extraction methods for optimal results. We also report retrieval performance for WiCE Figure \ref{fig:retrieval_perf} using measures nDCG@10 and Recall@10 using annotated documents provided for WiCE. We observe trends similar to overall task performance demonstrating that efficient retrieval approaches explored do not negatively impact task performance to a significant extent.

\noindent\textbf{Insight 2}: \textit{We find that extracting supporting facts improves efficiency across the entire pipeline, with Sparse setups achieving up to 2.0x speedups with only a minimal performance decline.}
\vspace{-1em}

\begin{table}[htb!]
\centering
\footnotesize
\begin{tabular}{l  c c c c }
\hline
\multirow{2}{*}{Method}  & \multicolumn{2}{c}{Total Latency}  & \multicolumn{2}{c}{Speedup} \\
\cline{2-5}
& CPU & GPU  &  CPU & GPU  \\ 
\hline \hline
\multicolumn{1}{l}{\colorg \textit{HoVer}} & \colorg & \colorg & \colorg & \colorg \\
 Full-Wiki (S+R) &  \multicolumn{2}{c}{659 ms} & - & - \\
Full-Wiki & 214  ms & 174 ms & 3.1x &  3.8x \\
% Original  &  55 ms & 13 ms  & - & 12 ms & 67 ms & 25 ms & 9.8x &  26.4x \\
Fact Extraction  & 60 ms & 21 ms & 11.0x & 31.4x  \\
Citation Extraction  & \textbf{51 ms} & \textbf{20 ms} & \textbf{\speedup{12.9x}} & \textbf{\speedup{33.0x}} \\
Fusion  & 63 ms & 24 ms &  10.5x & 27.5x \\
\hline
\multicolumn{1}{l}{\colorg\textit{WiCE}} & \colorg & \colorg & \colorg & \colorg   \\
 Full-Wiki (S+R) & \multicolumn{2}{c}{831  ms} & - & - \\
Full-Wiki &  292 ms & 238 ms & 2.8x  &  3.5x \\
% Original  &  95 ms & 43  ms  & - & 11 ms & 106 ms & 54 ms & 7.8x &  15.4x \\
Fact Extraction  & 103 ms & 48 ms & 8.1x & 17.3x \\
Citation Extraction  & 98 ms & 46 ms & 8.5x & 18.1x \\
Fusion  &\textbf{98 ms} & \textbf{46  ms} & \speedup{8.5x} &  \speedup{18.1x} \\
\hline
\end{tabular}
\caption{Latency and speedup measurements for Index compression setup. Speedup is compared with respect to the total latency of Sparse-retrieval + Re-rank (S+R) pipeline with the Full-Wiki setup. (S+R) runs on both CPU and GPU, sparse retrieval running on CPU and rest of components running on GPU}
%\vspace{-1cm}
\label{tab:jpq_latency}
\end{table}

%\input{tables/latency/jpq_latency}
% %%%%%%%%%%%%%%%%%%%%%%%%%%%%%%%%%%%%%%%%%%%%%%%%%%%%%%%%%%%%%%%%%%%%%%%%%%%%%%%%%%%
\vspace{-2em}
\subsection{Effectiveness of index compression on enhancing the efficiency of dense retrieval and fact-checking systems?}

To answer \textbf{RQ3}, we make use of index compression to further improve upon Dense Retrieval setups, not only with respect to memory requirements but also enhancing total inference latency compared to the sparse retrieve + re-rank setups in classical pipelines.  The index sizes of Wikipedia collection for standard dense retrieval are 7.51 GiB for HoVer and 9.70 GiB for WiCE. Using the JPQ index compression model with M=96 subvectors, we significantly reduced the storage space for vector embeddings from 1.5 KiB to 104.12 B. This reduced the HoVer index size to 544.89 MiB and the WiCE index to \textbf{672.95 MiB}, achieving a \textbf{93\% reduction (14.4:1 compression ratio)}. Further reducing subvectors could decrease the index size but may impact performance.

The utilization of JPQ index compression leads to significant reductions in retrieval latency compared to dense Retrieval and sparse retrieval, as demonstrated in \autoref{tab:jpq_latency}. CPU retrieval gains notable speedups of approximately 10.0x for HoVer experiments and 7.0x for WiCE experiments, while GPU retrieval shows about 2.0x and 0.8x speedups, respectively. These improvements are attributed to learned compression in JPQ, enhancing computational efficiency. 
Significant improvements are also observed when examining the inference latency of the whole pipeline. The CPU-based approaches shows impressive speedups (upto \textbf{12.9x} for HoVer and \textbf{8.5x} for WiCE), and GPU-based approaches achieve even higher gains (\textbf{33.0x} for HoVer and \textbf{18.1x} for WiCE). 

Surprisingly, in our experiments we observe that JPQ yields better results than standard Dense Retrieval as shown in Figure \ref{fig:performance_plot}. This is particularly due to joint training of the query encoder and index compression. In addition, JPQ employs end-end negative sampling, which further improves retrieval performance despite significant compression of embeddings.

\mpara{Insight 3}: \textit{We find that index compression reduces index size by \textbf{93\%} resulting in speedups for CPU-based setups up to 10x and GPU-based setups more than 20x compared to classical fact-checking pipeline.}

\subsection{Discussion of Live Fact-checking results}
\begin{figure}
\centering
 \hspace{6em}     \begin{subfigure}{.8\textwidth}
        \begin{tikzpicture}
\edef\mylst{"56.95","56.66","56.94",""}
\edef\explora{"55.92","57.82","52.93",""}

    \begin{axis}[
            ybar=14pt,
            width=6cm,
            bar width=0.35,
            every axis plot/.append style={fill},
            grid=major,
            xtick={1, 4, 8,9,11},
            xticklabels={Sparse + re-rank, Dense, IC},
            ylabel style = {font=\small},
        yticklabel style = {font=\boldmath \tiny,xshift=0.05ex},
        xticklabel style ={font=\tiny,yshift=0.5ex},
            ylabel={Performance (F1 weighted)},
            enlarge x limits=0.15,
            ymin=0,
            ymax=86,
            legend style ={font=\tiny,yshift=0.05ex},
            area legend,
            nodes near coords style={font=\tiny,align=center,text width=1em},
            legend entries={Full-Wiki, Fusion},
            legend cell align={left},
            legend pos=north west,
            legend columns=-1,
            legend style={/tikz/every even column/.append style={column sep=0.06cm}},
        ]
        \addplot+[
            ybar,
            plotColor1*,
            postaction={
                    pattern=north east lines
                },
                    nodes near coords=\pgfmathsetmacro{\mystring}{{\mylst}[\coordindex]}\textbf{\mystring},
        ] plot coordinates {
                (1,56.95)
                (4,56.66)
                (8,56.94)
            };
        \addplot+[
            ybar,
            plotColor2*,
            postaction={
                    pattern=north east lines
                },
            nodes near coords=\pgfmathsetmacro{\mystring}{{\explora}[\coordindex]}\textbf{\mystring},
        ] plot coordinates {
                (1,55.92)
                (4,57.82)
                (8,52.83)
                (9,0)
            };

    \end{axis}
\end{tikzpicture}

    \end{subfigure}
    \caption{Live fact-checking performance across different corpus setups}
    \label{fig:livefc}
\end{figure}
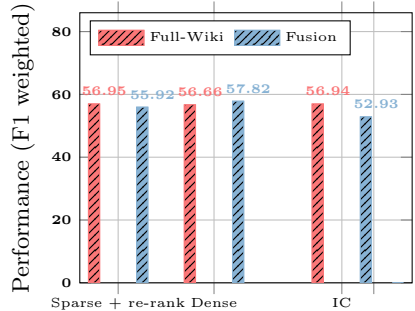

We employ the pruned index (2024 Wiki collection) using our Fusion approach followed by compression of the index for live Fact-checking of 2024 presidential debate. The pipeline comprises a dense retrieval using compressed index followed by claim verification. We use the query encoder and NLI models trained on HoVer for this application. We compare this approach to also the classical sparse-retrieval+re-rank fact-checking pipeline over the Full-Wiki collection (without pruning). The task performance is shown in Figure \ref{fig:livefc} and the corresponding pipeline efficiency is shown in Table \ref{tab:livefc}. We observe that the pruned collection coupled with retrieval using index compression leads to impressive speedups (\textbf{3.4x}) over classical pipeline over the full collection without significant drop in task performance (Figure \ref{fig:livefc}). The results demonstrate that efficient retrieval is critical for live fact-checking. Our experiments demonstrate that our approach for efficient retrieval provides significant speedups on CPUs further making the technology accessible even in low-resource scenarios which has significant implications in aiding detection of misinformation and disinformation at scale.
\vspace{-2em}
\begin{table}[htb!]
\centering
\footnotesize % Reduced font size
\setlength{\tabcolsep}{3pt} % Reduce space between columns
\renewcommand{\arraystretch}{0.9} % Reduce space between rows
\begin{tabular}{l c c c c c c}
\multirow{2}{*} & \multicolumn{2}{c}{\makecell{ Retrieval}}   & \multicolumn{2}{c}{Total Latency} & \multicolumn{2}{c}{Speedup} \\
\cline{2-3}\cline{4-5}\cline{5-7} \\[-1mm]
& CPU & GPU & CPU& GPU& CPU & GPU \\
\hline \\

% Term-based Document Retrieval
\colorg\textit{Sparse + Re-ranking} & \colorg & \colorg & \colorg  & \colorg & \colorg & \colorg\\ 
Full-Wiki & 463  & -   & 695  & - & \multicolumn{2}{c}{-} \\
Fusion & 274  & -   & 479   & - & \multicolumn{2}{c}{1.5x} \\
\hline \\
% Dense Retrieval setup
\colorg\textit{Dense Retrieval} & \colorg & \colorg & \colorg & \colorg & \colorg  & \colorg\\ 
Full-Wiki &  433   & 32  & 553   &  152  & 1.3x & 4.6x \\
Fusion & 412   & 32 & 511  & 131  & 1.4x & 5.3x \\
\hline \\
% Index Compression setup
\colorg\textit{Index Compression} & \colorg & \colorg & \colorg & \colorg & \colorg & \colorg\\  
Full-Wiki & 100  & 50   & 228  & 178  & 3.0x & 3.9x \\
\textbf{Fusion (ours)} & \textbf{89 }   & \textbf{43}  & \textbf{203}  & 157  & \speedup{3.4x} & 4.4x \\

\hline
\end{tabular}
\caption{Latency Comparisons for Live Fact-checking (in milliseconds (ms))}

\label{tab:livefc}
\end{table}

\vspace{0.5cm}
\section{Conclusion}
\label{conclusion}
%\todo {needs rewrite}
In this work, we assess how indexing potentially relevant facts from large data collections can improve fact-checking pipelines. Our experiments show that indexing only factual content coupled with index compression significantly improves pipeline efficiency with minimal performance loss, making it feasible for lower-end, CPU-focused machines and applications like live fact-checking. Future work will expand this approach by testing various settings for indexing, retrieval, and compression, using diverse claim datasets and larger corpora to strengthen the robustness and real-world applicability of fact-checking systems.
\section{Acknowledgements}
This work is partly funded by the Research Council of Norway project EXPLAIN (grant no:
337133). We acknowledge valuable
contributions from Factiverse AI team: Tobias Tykvart for Frontend, Henrik Vatndal  and
others for manual analysis (Maria Amelie,
Gaute Kokkvol, Sean Jacob, Christina Monets and
Mari Holand)

\newpage
\renewcommand*{\bibfont}{\scriptsize}
\bibliographystyle{abbrvnat}
\bibliography{reference}
%\clearpage
%\appendix
%\input{appendix}
\end{document}